\documentclass[aps,prl,twocolumn,epsfig,preprintnumbers,superscriptaddress,10pt]{revtex4}

\usepackage{graphicx}
\usepackage{dcolumn}
\usepackage{bm}

\newcommand{\be}{\begin{equation}}
\newcommand{\ee}{\end{equation}}
\newcommand{\bea}{\begin{eqnarray}}
\newcommand{\eea}{\end{eqnarray}}
\newcommand{\ba}{\begin{eqnarray}}
\newcommand{\ea}{\end{eqnarray}}
\newcommand{\nn}{\nonumber \\}
\newcommand{\eqn}[1]{(\ref{#1})}
\newcommand{\beq}{\begin{equation}}
\newcommand{\eeq}{\end{equation}}
\newcommand{\beqa}{\begin{eqnarray}}
\newcommand{\eeqa}{\end{eqnarray}}
\newcommand{\beqar}{\begin{eqnarray*}}
\newcommand{\eeqar}{\end{eqnarray*}}

\newcommand{\cala}{{\cal A}}
\newcommand{\calj}{{\cal J}}

\newcommand{\vlim}{v_\mt{lim}}

\def\nc {N_\mt{c}}
\def\nf {N_\mt{f}}

\def\t7 {T_\mt{D7}}

\newcommand{\mq}{M_\mt{q}}      

\newcommand{\mt}[1]{\textrm{\tiny #1}}

\begin{document}


\title{A New Mechanism of Quark Energy Loss}
\author{Jorge Casalderrey-Solana} 
\affiliation{Physics Department, Theory Unit, CERN, CH-1211 Gen\`eve 23, Switzerland}
\author{Daniel Fern\'andez}
\affiliation{Departament de F\'\i sica Fonamental \&  Institut de Ci\`encies del Cosmos, Universitat de Barcelona, Diagonal 647, E-08028 Barcelona, Spain}
\author{David Mateos$^{2,}$}
\affiliation{Instituci\'o Catalana de Recerca i Estudis Avan\c cats (ICREA), Llu\'\i s Companys 23, E-08010, Barcelona, Spain}

\preprint{CERN-PH-TH/2009-262}
\preprint{ICCUB-09-440}


\begin{abstract}
We show that a heavy quark moving sufficiently fast through a quark-gluon plasma may lose energy by Cherenkov-radiating mesons. We demonstrate that this takes place in all strongly coupled, large-$\nc$ plasmas with a gravity dual. The energy loss is exactly calculable in these models despite being an ${\cal O}(1/\nc)$-effect. We discuss phenomenological implications for heavy-ion collision experiments. 
\end{abstract}

\maketitle

\noindent {\bf 1. Introduction.}
A remarkable conclusion from the Relativistic Heavy Ion Collider (RHIC) experiments \cite{rhic} is that the quark-gluon plasma does not behave as a weakly coupled gas of quarks and gluons, but rather as a strongly coupled fluid \cite{fluid}. This makes the study of the plasma a challenging task.

Experimentally, valuable information is obtained by analyzing the energy loss of energetic partons created in hard initial collisions. In order to use this information to learn about the plasma, a theoretical, quantitative understanding of the different mechanisms of parton energy loss is needed. Several such mechanisms have been previously studied, both in QCD itself \cite{radiative} and in the context of the gauge/gravity duality \cite{strong}.

In this letter we will uncover a new mechanism whereby a sufficiently fast heavy quark traversing a strongly coupled plasma loses energy by Cherenkov-radiating in-medium mesons. We will first show that this takes place in all strongly coupled, large-$\nc$ theories with a gravity dual. Next we will calculate the energy loss in a simple example. Finally, we will discuss possible implications for heavy-ion collision experiments.

\begin{figure}
\includegraphics[scale=.42]{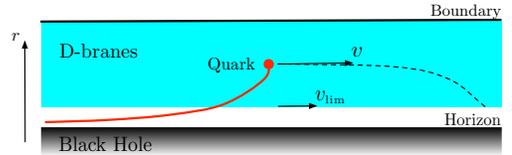}
\caption{D-branes and open string in a BH geometry.} 
\label{cherenkov-with-quark}
\end{figure}

\noindent
{\bf 2. Universality of the mechanism.}
This follows from two universal properties of the gauge/gravity duality (in the limit $\nc, \lambda \rightarrow \infty$): (i) the fact that the gauge theory deconfined phase is described by a black hole (BH) geometry \cite{Witten}, and (ii) the fact that a finite number of quark flavours $\nf$ is described by $\nf$ D-brane probes \cite{Karch-Randall} -- see fig.~\ref{cherenkov-with-quark}. 
In addition to the gauge theory directions, the gravity description always includes a radial direction which is dual to the gauge theory energy scale. The radial position of the horizon is proportional to the plasma temperature $T$. The D-branes extend in the radial direction down to a minimum value proportional to the (constituent) quark mass $\mq$. 

For sufficiently large $\mq/T$, the D-branes sit outside the horizon \cite{MMT,dpdq, dpdqdqbar}. In this phase, low-spin gauge theory mesons are described by small, normalizable fluctuations of scalar and vector fields propagating on the branes,
whose spectrum is discrete and gapped. In particular, this means that sufficiently heavy mesons survive deconfinement, in agreement with lattice and potential model predictions for real-world QCD \cite{quarkonium}.

\begin{figure}
\includegraphics[scale=.55]{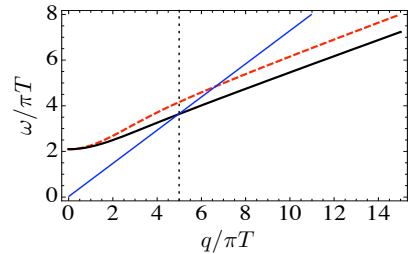}
\caption{DR for the transverse (black, continuous curve) and longitudinal (red, dashed curve) modes of a heavy vector meson with $\vlim = 0.35$ in the D3/D7 system. The blue, continuous straight line corresponds to $\omega = v q$ with $\vlim<v<1$.} 
\label{dispersion}
\end{figure}

Let $\omega(q)$ be the in-medium dispersion relation (DR) for these mesons.
As an illustrative example, the DR for vector mesons in the D3/D7 system is depicted in fig.~\ref{dispersion}. 
As $q \rightarrow \infty$, the DR becomes linear: $\omega(q) \sim \vlim q$, with $\vlim < 1$. This subluminal limiting velocity, which is the same for all mesons, is easy to understand in the gravitational description \cite{limiting}. Since highly energetic mesons are strongly attracted by the BH, their wave-function is very concentrated at the bottom of the branes. Consequently, their velocity is limited by the local speed of light $\vlim$ at this point (see fig.~\ref{cherenkov-with-quark}). Because of the BH redshift, $\vlim$ is lower than the speed of light at infinity. In the gauge theory this translates into the statement that $\vlim$ is lower than the speed of light in the vacuum \cite{MMT2}.  

Consider now a heavy quark in the plasma. In the gravitational picture, this is described by a string that starts on the D-branes and falls through the horizon -- see fig.~\ref{cherenkov-with-quark}. 
In order to model a highly energetic quark we consider a string whose endpoint moves with an arbitrary velocity $v$ at an arbitrary radial position $r_0$, where $r_0$ is inversely proportional to the size of the gluon cloud that dresses the quark \cite{size}.

Two simple observations now lead to the effect that we are interested in. The first one is that the string endpoint is charged under the scalar and vector fields on the branes. In the gauge theory, this corresponds to an effective quark-meson coupling (see fig. 3) of order $\sim 1/\sqrt{\nc}$. 
\begin{figure}
\includegraphics[scale=.30]{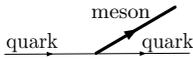}
\caption{Effective quark-meson coupling.} 
\label{quark-meson-coupling}
\end{figure}
The dynamics of the branes+string endpoint system is thus (a generalization of) that of classical electrodynamics in a medium in the presence of a fast-moving charge. The second observation is that the velocity of the quark may exceed the limiting velocity of the mesons, since the redshift at the position of the string endpoint is smaller than at the bottom of the branes. As in ordinary electrodynamics, if this happens then the string endpoint loses energy by Cherenkov-radiating into the fields on the brane. In the gauge theory, this translates into the quark losing energy by Cherenkov-radiating scalar and vector mesons. The rate of energy loss is set by the square of the coupling, and is therefore of order $1/\nc$.

\noindent
{\bf 3. A quantitative example.}
In this section we will calculate the rate of energy loss in four-dimensional, $SU(\nc)$, 
${\cal N}=4$ super Yang-Mills (SYM) theory coupled to one quark flavour (but the result is valid for arbitrary $\nf$, see sec.~5). The dual description consists of a D7-brane probe in the supergravity background of $\nc$ black D3-branes. Following \cite{MMT2} we write the induced metric on the D7-brane worldvolume as $ds^2 = L^2 ds^2 (g)$ with
\be
ds^2 (g) =  \frac{\rho^2}{2} \left[-\frac{f^2}{\tilde f} d t^2 + {\tilde f} dx_i^2\right]
+ \frac{(1+\dot R ^2)}{\rho^2}  dr^2+ \frac{r^2}{\rho^2} d \Omega^2_3 \,,
\label{g}
\ee
where $\rho^2 = R^2 + r^2$, $f = 1 - 1/\rho^4$, $\tilde{f} = 1 + 1/\rho^4$, $\dot{R}=dR/dr$, and 
$x^\mu=\{t, x^i\}$ are the four gauge theory directions. $R(r)$ describes the D7-brane embedding, with $R(\infty)=2\mq/\sqrt{\lambda}T$. The  dimensionless coordinates above are related to their dimensionful counterparts (denoted with tildes) through $x^\mu = \pi T \, \tilde{x}^\mu$, 
$\{r, R, \rho \} = \{\tilde{r}, \tilde{R}, \tilde{\rho} \}/\pi L^2 T$.

The terms in the brane+string action relevant to our calculation are
\be
S= -  \int d^8 \sigma \sqrt{-g}\, \frac{1}{4} F^{ab}F_{ab}   
-  e \int d\tau A_a \frac{d\sigma^a}{d\tau} \,,
\label{8action}
\ee
where $F_{ab}=\partial_{[a} A_{b]}$ and $\sigma^a=\{ x^\mu, r, \Omega_3 \}$. The first term comes from expanding the Dirac-Born-Infeld part of the D7-brane action to quadratic order in the gauge field. The metric $g$ that enters this term is that in eqn.~\eqn{g}, which contains no factors of $L$; these have been absorbed in the definition of $e$ in eqn.~\eqn{e}. The Wess-Zumino part of the D7-brane action will not contribute to our calculation. The second term in \eqn{8action} is the minimal coupling of the endpoint of an open string, whose worldline is parametrised by 
$\sigma^a(\tau)$, to the gauge field on the branes. We have omitted a similar coupling to the scalar fields, which will be considered in \cite{cherenkov}. The coupling constant in \eqn{8action} is
\be
e^2= \frac{1}{T_\mt{D7} \left(2 \pi l^2_s\right)^2 L^4} = \frac{8 \pi^4}{\nc} \,,
\label{e}
\ee
where $T_\mt{D7} = 1/g_s (2\pi)^7 \ell_s^8$ is the D7-brane tension. As expected, $e$ is of order $1/\sqrt{\nc}$, which justifies our neglect of terms of order higher than quadratic in the action. 

The second term in \eqn{8action} may be written as $-e \int d^8\sigma A_a J^a$. For simplicity, we will assume that the quark moves with constant velocity along a straight line at constant radial and angular positions, so we write
\be
J^a = \delta^{(3)}( \vec{x} - \vec{v} t ) \, \delta (r - r_0) \, \delta^{(3)}( \Omega - \Omega_0 ) \times (1, \vec{v}, 0, \vec{0} )\,.
\ee
In reality, $r_0$ and $v$ will of course decrease with time because of the BH gravitational pull and the energy loss. However, for simplicity we will concentrate on the initial part of the trajectory (which is long provided the initial quark energy is large) for which $r_0$ and $v$ are approximately constant \cite{trajectory} 
-- see fig.~\ref{cherenkov-with-quark}.

The rate of quark energy loss is given by minus the work per unit time done by the gauge field:
\be
\frac{dE}{dt} = -e \int d^3 x dr d\Omega_3 \, F_{0a} J^a 
= - e v^i F_{0i}(t, \vec{v} t, r_0, \Omega_0) \,.
\label{loss}
\ee 
Since real-world QCD has no internal $S^3$, we focus on modes with no angular momentum on the $S^3$. These take the form $A_\mu(x^\nu, r), A_r(x^\nu, r), A_\Omega=0$. We set $A_r=0$ by a gauge choice. Further, we work with the Fourier-space components $A_\mu(\omega,q,r)$ and choose $\vec{q}=(q,0,0)$, $\vec{v}=v(\cos\theta, v\sin \theta, 0)$. After integrating over the $S^3$, the relevant Fourier-space components of the current are 
\be
J^\mu = 2\pi \delta( \omega - q v \cos \theta ) \, \delta (r - r_0) \times (1,\cos \theta, \sin \theta, 0) \,.
\label{J}
\ee
With this choice the only transverse mode of the gauge field excited by the source is 
$\cala = A_2$. The equation of motion for this mode is
\be
\partial_r \left( \frac{f r^3 \, \partial_r \cala}{2\sqrt{1+  \dot R ^2}}  \right)
+  \sqrt{1+\dot R^2} \frac{r^3}{\rho^4} \left(\frac{\omega^2 \tilde{f}}{f}
- \frac{q^2 f}{\tilde{f}} \right) \cala = {\tilde e} \calj \,, 
\label{trans}
\ee
where $\calj = J^2$, $\tilde{e} = e/\Omega_3$ and $\Omega_3=2\pi^2$ is the volume of a unit $S^3$. We solve \eqn{trans} by expanding $\cala$ as
\be
\cala (\omega, q, r) = \sum_n \cala_n (\omega, q) \, \xi_n (q,r)
\label{expansion}
\ee
in terms of a basis of normalizable eigen-functions $\{ \xi_n (q,r) \}$ in the radial direction. These are solutions of eqn.~\eqn{trans} with $\calj=0$ with $q$-dependent eigen-values $\omega=\omega_n(q)$, and satisfy the orthonormality relations 
\be
\int_0^\infty dr \frac{\tilde{f} r^3}{f \rho^4} \, \sqrt{1+\dot R^2}  \, \xi_m \xi_n 
=\delta_{mn} \,.
\ee
Inserting the expansion \eqn{expansion} in \eqn{trans}, and using the eigen-state equation and the orthonormality relations, we find 
\be
\left[ \omega^2 - \omega_n^2(q) \right]  \cala_n(\omega,q) 
= {\tilde e} \calj_n (\omega, q)\,,
\label{fourdim}
\ee
where
\bea
\calj_n (\omega, q) &=& \int dr \calj(\omega, q) \xi_n (q,r) \nonumber \\
&=& 2\pi \delta\left(\omega- q v \cos\theta \right) v \sin \theta \, \xi_n (q,r_0) \,.
\label{Jn}
\eea
Through the expansion \eqn{expansion} we have `Kaluza-Klein' reduced the five-dimensional gauge field to a discrete, infinite tower of independent four-dimensional gauge fields 
$\{ \cala_n(\omega,q)\}$. Each of these fields is characterized by a $q$-dependent radial `wave-function' $\xi_n (q,r)$, as well as by a DR $\omega=\omega_n(q)$, and couples to the quark with an effective strength $e_\mt{eff}(q,r_0)=e \xi_n (q,r_0)$. 

\begin{figure}
\includegraphics[scale=.36]{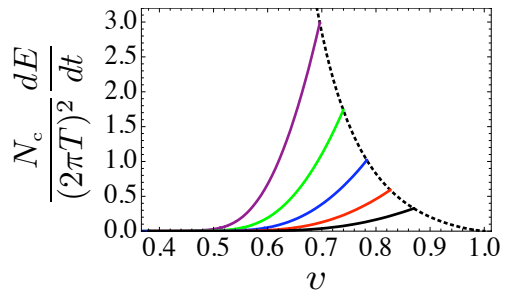}
\caption{Energy loss into ${\cal A}_{(n=0)}$ for an embedding with $R(\infty)=1.32$. The continuous curves correspond to $r_0=0.86, 0.97, 1.10, 1.25, 1.45$. The dotted curve is defined by the endpoints of the constant-$r_0$ curves.} 
\label{result}
\end{figure}

With retarded boundary conditions, as appropriate for the reaction to the quark's passage, eqn.~\eqn{fourdim} yields 
\be
\cala_n(\omega,q) = 
\frac{\tilde{e} \calj_n (\omega, q)}{\left( \omega + i\epsilon \right)^2 - \omega_n^2(q)} \,.
\ee  
We now evaluate \eqn{loss} to obtain the energy deposited on the $n$-th transverse mode. We first express $F_{02}(t, \vec{v} t, r_0, \Omega_0)$ as an integral over its Fourier components. We then integrate over frequencies trivially because of the delta-function in \eqn{Jn}. Finally, we set $d^3 q= 2\pi q^2 dq ds$, where $s=\cos \theta$, to arrive at
\bea
\frac{dE_{n}}{dt} &=& -\frac{e^2 v}{\Omega_3} \int_0^\infty \frac{dq}{2\pi} \,
q \xi_n^2 (q,r_0) \int_{-1}^1 \frac{ds}{2\pi i} 
\frac{s(1-s^2)}{(s+i\epsilon)^2 - s_n^2(q)} \nn
&=& \frac{e^2 v}{2\Omega_3} \int_0^\infty \frac{dq}{2\pi} \,
q \xi_n^2 (q,r_0) (1 - s_n^2(q)) \Theta (1 - s_n^2(q)) \,, \nn
\label{losstrans}
\eea 
where $s_n = v_n/v$ and $v_n (q) = \omega_n (q)/q$ is the phase velocity of the mode. The Heaviside function confirms the expected result: The quark only radiates into modes with phase velocity lower than $v$ -- those to the right of the dashed, vertical line in  fig.~\ref{dispersion}. 
The numerical result for $n=0$ is plotted in fig.~\ref{result}.
For fixed $r_0$, the energy loss increases monotonically with $v$ up to the maximum allowed value of $v$ -- the local speed of light at $r_0$. As $r_0$ decreases, the characteristic momentum $q_\mt{char}$ of the modes contributing to the integral increases. 
As $r_0 \rightarrow 0$ these modes become increasingly peaked at small $r$, and $e_\mt{eff}(q_\mt{char},r_0)$ and the energy loss diverge  \cite{cherenkov}. However, this mathematical divergence is removed by physical effects we have not taken into account. For example, for sufficiently large $q$ the radial profile of the mesons becomes of order the string length and stringy effects become important \cite{mit}. Also, mesons acquire widths $\Gamma \propto q^2$ at large $q$ \cite{width} and can no longer be treated as well defined quasiparticles. Finally,
the approximation of a constant-$v$, constant-$r_0$ trajectory ceases to be valid whenever the energy loss rate becomes large. 

\noindent
{\bf 4. Phenomenology.}
The Cherenkov radiation of mesons by quarks depends only on the qualitative features of the DR of fig.~\ref{dispersion}, which are universal for all gauge theory plasmas with a dual gravity description \cite{implications}. Moreover, it is conceivable that they may also hold for QCD mesons such as the $J/\psi$ or the $\Upsilon$ (see {\it e.g.} the discussion in \cite{peak}). Here we will examine some qualitative consequences of this assumption for HIC experiments.

The energy lost into mesons would be reflected in a reduction of the heavy quark nuclear modification factor $R_\mt{AA}$ \cite{light}. This would only occur for high enough quark velocities, thus yielding a very particular behaviour of $R_\mt{AA}$. Note that the minimum quark velocity at which the reduction starts to occur may actually be higher than $\vlim$, since the quark energy must be larger than the mass of the radiated meson. For example, for a charm quark to radiate a $J/\psi$ meson this condition yields $v > 0.87$. In fact, our calculation applies strictly only in the limit of infinite quark energy, which suggests that it should be more relevant to HIC experiments at the Large Hadron Collider than at RHIC. 

The radiated mesons would be preferentially emitted at a characteristic Cherenkov angle 
$\cos \theta_\mt{c}=\vlim/v$. Taking the gravity result as guidance, $\vlim$ could be as low as $\vlim=0.35$ at the meson dissociation temperature \cite{MMT2}, corresponding to an angle as large as 
$\theta_\mt{c} \approx 1.21$ rad. This emission pattern is similar to the emission of sound waves by an energetic parton \cite{mach} in that both effects lead to a non-trivial angular structure. One important difference, however, is that the radiated heavy mesons would not thermalize and hence would not be part of a hydrodynamic shock wave. As in the Mach cone case, the meson emission pattern could be reflected in azimuthal dihadron correlations triggered by a high-$p_\mt{T}$ hadron. Due to surface bias, the energetic parton in the triggered direction is hardly modified, while the one propagating in the opposite direction moves through a significant amount of medium, emitting heavy mesons. Thus, under the above assumptions, the dihadron distribution with an  associated $J/\psi$ would have a ring-like structure  peaked at an angle 
$\theta \approx \pi-\theta_\mt{c}$.

\noindent
{\bf 5. Discussion.} 
Cherenkov emission of gluons in the context of heavy ion collisions has been considered in \cite{gluon}, where the in-medium gluons are assumed to have space-like DR. Although some of the underlying physics is similar, the mechanism we have discussed is different in two respects. First, the radiated particles are colourless mesons, not gluons. Second, the gauge/gravity duality provides a large class of completely explicit and calculable examples in which this mechanism is realized. 

We have focused on the transverse modes of the gauge field. Since the vector mesons are massive, there is a similar energy loss into the longitudinal modes \cite{cherenkov}. 

We calculated the energy deposited on the branes by the string endpoint. Since the branes sit outside the BH, this energy must stay on the branes (in the limit $\nc, \lambda \rightarrow \infty$). Because total conserved charges must agree, this energy is the same as the energy lost by the quark in the gauge theory. The Cherenkov angle $\theta_\mt{c}$ is also the same, since it is determined by kinematics alone. In contrast, extracting unintegrated gauge theory quantities ({\it e.g.} the differential power spectrum) would require computing the boundary stress-energy tensor as in \cite{boundary}.

The validity of our results requires not just large $\nc$, but also strong coupling. In this regime the holographic mesons behave as elementary excitations, as opposed to composite bound states, up to energies of order $\sqrt{\lambda} T \gg T$ \cite{MMT,MMT2}. Despite the strong coupling requirement, our results might apply to asymptotically free theories as long as:  (i) they are sufficiently strongly coupled at the scale $T$, so that the in-medium meson DR shares the qualitative features of that in fig.~\ref{dispersion}, and (ii) there is a non-zero quark-meson coupling in the medium. 

The energy loss is $\nf$-independent because (at leading order) the string endpoint couples directly to the gauge field on only one of the $\nf$ D-branes. The  $1/\nc$ scaling of the energy loss does not necessarily imply that the analogous effect (if present) is small in $\nc=3$ real-world QCD, in particular at high quark velocities. Furthermore, its characteristic geometry and velocity dependence may make it easily identifiable.

We close with a comment on the energy loss of heavy mesons. At $\nc \rightarrow \infty$, these mesons experience no drag \cite{no-drag, MMT2}. 
At finite $\nc$, pointlike heavy mesons experience a drag of order ${\cal O}(1/\nc^2)$ \cite{pointlike}. Cherenkov radiation implies an ${\cal O}(1/\nc)$ drag for fast excited mesons describable as a long string with both endpoints on the D-branes \cite{higher}, since each endpoint may radiate as an individual quark. 

\noindent 
{\bf Acknowledgments.} We thank M.\ Chernicoff, R.~Emparan, B.~Fiol, A.~Guijosa, C.\ Manuel, A.~Paredes, L.~Pati\~{n}o and P.~Townsend for discussions. We are supported by a Marie Curie
PIEF-GA-2008-220207 (JCS) and 2009SGR168, MEC FPA 2007-66665-C02 and CPAN CSD2007-00042 Consolider-Ingenio 2010 (DF, DM).


\begin{thebibliography}{99}

\bibitem{rhic}
J.~Adams {\it et al.}  [STAR Collaboration],
Nucl.\ Phys.\  A {\bf 757}, 102 (2005);
%
%
K.~Adcox {\it et al.}  [PHENIX Collaboration],
Nucl.\ Phys.\  A {\bf 757}, 184 (2005).

\bibitem{fluid}
  E.~Shuryak,
  Prog.\ Part.\ Nucl.\ Phys.\  {\bf 53}, 273 (2004);
%
%
  Nucl.\ Phys.\  A {\bf 750}, 64 (2005).

\bibitem{radiative}
  M.~Gyulassy {\it et al}, 
  arXiv:nucl-th/0302077;
  A.~Kovner and U.~A.~Wiedemann,
  arXiv:hep-ph/0304151;
  J.~Casalderrey-Solana and C.~A.~Salgado,
  Acta Phys.\ Polon.\  B {\bf 38}, 3731 (2007).

\bibitem{strong}
  C.~P.~Herzog {\it et al}, 
  JHEP {\bf 0607}, 013 (2006);
%
  H.~Liu {\it et al}, 
  Phys.\ Rev.\ Lett.\  {\bf 97}, 182301 (2006);
  S.~S.~Gubser,
  Phys.\ Rev.\  D {\bf 74}, 126005 (2006);
%
  J.~Casalderrey-Solana and D.~Teaney,
  Phys.\ Rev.\  D {\bf 74}, 085012 (2006);
  M.~Chernicoff and A.~Guijosa,
  JHEP {\bf 0806}, 005 (2008).

\bibitem{Witten}
  E.~Witten,
  Adv.\ Theor.\ Math.\ Phys.\  {\bf 2}, 505 (1998).

\bibitem{Karch-Randall}
  A.~Karch and L.~Randall,
  JHEP {\bf 0106}, 063 (2001);
  A.~Karch and E.~Katz,
  JHEP {\bf 0206}, 043 (2002).


\bibitem{MMT}
  D.~Mateos {\it et al}, 
  Phys.\ Rev.\ Lett.\  {\bf 97}, 091601 (2006).

\bibitem{dpdq}
  J.~Babington {\it et al}, 
  Phys.\ Rev.\  D {\bf 69}, 066007 (2004);
  I.~Kirsch,
  Fortsch.\ Phys.\  {\bf 52}, 727 (2004);
  M.~Kruczenski {\it et al}, 
  JHEP {\bf 0405}, 041 (2004).

\bibitem{dpdqdqbar}
  O.~Aharony {\it et al}, 
  Annals Phys.\  {\bf 322}, 1420 (2007);
  A.~Parnachev and D.~A.~Sahakyan,
  Phys.\ Rev.\ Lett.\  {\bf 97}, 111601 (2006).

\bibitem{quarkonium}
See {\it e.g.} A.~Bazavov {\it et al}, 
  arXiv:0904.1748 [hep-ph], and references therein. 

\bibitem{limiting}
This was first noted in \cite{MMT2} and subsequently elaborated upon in \cite{mit}.

\bibitem{MMT2}
D.~Mateos {\it et al}, 
JHEP {\bf 0705}, 067 (2007).

\bibitem{mit}
  Q.~J.~Ejaz {\it et al}, 
  JHEP {\bf 0804}, 089 (2008).

\bibitem{size}
  J.~L.~Hovdebo {\it et al}, 
  Int.\ J.\ Mod.\ Phys.\  A {\bf 20}, 3428 (2005).

\bibitem{cherenkov}
J.~Casalderrey-Solana, D.~Fern\'andez and D.~Mateos, in preparation.

\bibitem{trajectory}
  P.~M.~Chesler {\it et al}, 
  Phys.\ Rev.\  D {\bf 79}, 125015 (2009).

\bibitem{implications}
Implications for photon production have been discussed in \cite{peak}, and for deep inelastic scattering in \cite{DIS}.

\bibitem{peak}
  J.~Casalderrey-Solana and D.~Mateos,
  Phys.\ Rev.\ Lett.\  {\bf 102}, 192302 (2009).

\bibitem{DIS}
  E.~Iancu and A.~H.~Mueller,
  arXiv:0912.2238 [hep-th].
  
\bibitem{width}
  T.~Faulkner and H.~Liu,
  Phys.\ Lett.\  B {\bf 673}, 161 (2009).
  
\bibitem{light}
If light quarks can also radiate, the light hadron $R_\mt{AA}$ would also show this behaviour.
  
\bibitem{gluon}
  V.~Koch {\it et al}, 
  Phys.\ Rev.\ Lett.\  {\bf 96}, 172302 (2006);
  I.~M.~Dremin,
  Nucl.\ Phys.\  A {\bf 767}, 233 (2006).

\bibitem{mach}
  J.~Casalderrey-Solana {\it et al}, 
  J.\ Phys.\ Conf.\ Ser.\  {\bf 27}, 22 (2005)
  [Nucl.\ Phys.\  A {\bf 774}, 577 (2006)].

\bibitem{boundary}
  J.~J.~Friess {\it et al},
  Phys.\ Rev.\  D {\bf 75}, 106003 (2007);
  P.~M.~Chesler and L.~G.~Yaffe,
  Phys.\ Rev.\  D {\bf 78}, 045013 (2008).

\bibitem{no-drag}
  K.~Peeters {\it et al}, 
  Phys.\ Rev.\  D {\bf 74}, 106008 (2006);
  H.~Liu {\it et al}, 
  Phys.\ Rev.\ Lett.\  {\bf 98}, 182301 (2007);
  M.~Chernicoff {\it et al}, 
  JHEP {\bf 0609}, 068 (2006).

\bibitem{pointlike}
  K.~Dusling {\it et al}, 
  JHEP {\bf 0810}, 098 (2008).

\bibitem{higher}
Attached at points higher than the bottom of the branes.

\end{thebibliography}
\end{document}